\documentclass[aps,prr,twocolumn,superscriptaddress,floatfix,nopacs,longbibliography]{revtex4-1}
\usepackage{graphicx,amsfonts,amssymb,amsmath,hyperref,enumerate}

\newif\ifhyper
\hypertrue
\ifhyper
\hypersetup{
   citecolor = {green},
   colorlinks = {true}, 
   urlcolor = {blue} 
}
\fi

\newcommand{\beq}{\begin{equation}}
\newcommand{\eeq}{\end{equation}}
\newcommand{\beqa}{\begin{eqnarray}}
\newcommand{\eeqa}{\end{eqnarray}}
\newcommand{\ket} [1] {\vert #1 \rangle}

\newcommand{\widebar}[1]{\overline{#1}}

\newcommand{\ro}[1]{{\color{black}{#1}}}

\newcommand{\w}{\omega}

\def\ket#1{\vert#1\rangle}

\def\Longarrow{\protect\@lra}
\def\@lra{\relbar\joinrel\relbar\joinrel\relbar\joinrel%
          \relbar\joinrel\rightarrow}

\begin{document}

\title{Dynamic Portfolio Optimization with Real Datasets Using \\ Quantum Processors and Quantum-Inspired Tensor Networks}

\author{Samuel Mugel}
\affiliation{Multiverse Computing, Banting Institute, 100 College Street, ONRamp Suite 150, Toronto, ON M5G 1L5 Canada}

\author{Carlos Kuchkovsky}
\affiliation{BBVA Research \& Patents, Calle Sauceda 28, 28050 Madrid, Spain}

\author{Escol\'astico S\'anchez}
\affiliation{BBVA Research \& Patents, Calle Sauceda 28, 28050 Madrid, Spain}

\author{Samuel Fern\'andez-Lorenzo}
\affiliation{BBVA Research \& Patents, Calle Sauceda 28, 28050 Madrid, Spain}

\author{Jorge Luis-Hita}
\affiliation{BBVA Research \& Patents, Calle Sauceda 28, 28050 Madrid, Spain}

\author{Enrique Lizaso} 
\affiliation{Multiverse Computing, Paseo de Miram\'on 170, E-20014 San Sebasti\'an, Spain}

\author{Rom\'an Or\'us}
\affiliation{Multiverse Computing, Paseo de Miram\'on 170, E-20014 San Sebasti\'an, Spain}
\affiliation{Donostia International Physics Center, Paseo Manuel de Lardizabal 4, E-20018 San Sebasti\'an, Spain}
\affiliation{Ikerbasque Foundation for Science, Maria Diaz de Haro 3, E-48013 Bilbao, Spain}

\begin{abstract}
In this paper we tackle the problem of dynamic portfolio optimization, i.e., determining the optimal trading trajectory for an investment portfolio of assets over a period of time, taking into account transaction costs and other possible constraints. This problem is central to quantitative finance. After a detailed introduction to the problem, we implement a number of quantum and quantum-inspired algorithms on different hardware platforms to solve its discrete formulation using real data from daily prices over 8 years of 52 assets, and do a detailed comparison of the obtained Sharpe ratios, profits and computing times. In particular, we implement classical solvers (Gekko, exhaustive), D-Wave Hybrid quantum annealing, two different approaches based on Variational Quantum Eigensolvers on IBM-Q (one of them brand-new and tailored to the problem), and for the first time in this context also a quantum-inspired optimizer based on Tensor Networks. In order to fit the data into each specific hardware platform, we also consider doing a preprocessing based on clustering of assets. From our comparison, we conclude that D-Wave Hybrid and Tensor Networks are able to handle the largest systems, where we do calculations up to 1272 fully-connected qubits for demonstrative purposes. Finally, we also discuss how to mathematically implement other possible real-life constraints, as well as several ideas to further improve the performance of the studied methods.

\end{abstract}

\maketitle

\section{Introduction}
\label{intro}

In quantitative finance, portfolio optimization is the problem of selecting the best distribution of assets that optimizes some objective function \cite{portfolio}. Typically, this objective function tries to maximize the expected returns and minimize the financial risk. The problem gets more complicated if we do it dynamically, i.e., optimize the investment portfolio over a series of consecutive trading days. In this \emph{dynamic portfolio optimization}, the goal is to determine the optimal trading trajectory over the considered period of time, i.e., the optimal decisions that should be taken (or should have been taken) by a broker in order to maximize the overall return at the end of the time period. The dynamic problem is more complex because transactions' costs and transactions' market impact must be taken into account, as well as other possible constraints. In practice, it is well-known that this is an \emph{intractable} problem \ro{for generic instances}. 

Parallel to the above, it has been understood recently that quantum and quantum-inspired computing can help in solving hard financial problems \cite{qcreview, qcreview2}. For instance, a quantum computer should be able to solve more efficiently and with more accuracy problems related to pricing of financial derivatives \cite{Rebentrost2018, derivatives1, derivatives2}, prediction of financial crashes \cite{crash1, crash2}, detection of arbitrage cycles \cite{Rosenberg}, credit scoring \cite{Milne}, and identification of several types of fraud, among many applications. Portfolio optimization is no exception, as observed in several preexisting studies \cite{Rosenberg2016, delprado}. Yet, to the best of our knowledge, none of these works aimed to solve the problem on real datasets, and no comparison has ever been done openly and democratically between different methods and hardware platforms. 

In this paper we implement several quantum and quantum-inspired algorithms for dynamic portfolio optimization, and run them for the first time (as fas as we are aware) with real data corresponding to daily prices over 8 years of 52 assets. In particular, we implement D-Wave Hybrid quantum annealing, a Variational Quantum Eigensolver (VQE) on a quantum processor of IBM-Q, a new VQE-inspired algorithm which we call ``VQE Constrained" also on IBM-Q, and a quantum-inspired Tensor Network (TN) optimization algorithm (which is also the first implementation of a TN algorithm to solve a real and practical financial problem). We benchmark our algorithms using two classical methods: a Gekko solver (a Python-based optimization suite \cite{Beal2018}) and an exhaustive solver. We expose how preprocessing is performed, and reduce the problem's dimensionality using a clustering algorithm. We then do a detailed comparison of all the results, focusing on the obtained Sharpe ratios and computing times.  From our comparison we conclude that, as of today, D-Wave Hybrid and Tensor Networks are able to handle the largest systems, where we do calculations up to 1272 fully-connected qubits for demonstrative purposes. Interestingly, we see that there is no clear answer as to which is the ``best" algorithm and/or hardware platform to deal with large systems, as this depends strongly on different figures of merit. 

This paper is organized as follows. In Sec.\,\ref{problem} we give an overview of the dynamic portfolio optimization problem. We show how it can be expressed Quadratic Unconstrained Binary Optimization (QUBO) problem, and discuss the differences between the continuous and discrete formulations. In Sec.\,\ref{tech} we give a very brief overview of the D-Wave hybrid, VQE, and TN algorithms. In Sec.\, \ref{data} we expose the data-preparation procedure, which reduces the problem's dimensionality by identifying irrelevant assets and performing clusterization. Sec.\,\ref{results} compares results obtained using the Gekko, Exhaustive, D-Wave Hybrid, VQE, VQE-Constrained, and TN solvers. Sec.\,\ref{next} discusses industry relevant next steps, such as the inclusion of more real-life constraints (e..g, market impact, exact linear transaction costs, and the so-called $10-5-40$ rule) and potential performance improvements. Finally, in Sec.\,\ref{conclusion} we wrap up and conclude.  

\section{Problem overview}
\label{problem}

The dynamic portfolio optimization problem can be expressed in a form amenable to a quantum computer. In what follows, we present the problem in some technical detail. Where possible, we use the notation of \cite{Rosenberg2016}.

\subsection{Optimal dynamic portfolio}

\subsubsection{Generalities} 

In the dynamic version of the so-called \emph{Modern Portfolio Theory} (or \emph{Mean Variance Analysis}), we deal with the issue of allocating weigths to a number of assets over a period of time, in order to maximize the overall return at the end of the period. More specifically, for $N$ assets we consider an $N$-dimensional vector of weigths $\widebar{\w}_{t}$. Each of its component $\widebar{\w}_{n,t}$ is the weigth of asset $n$ at time $t = t_i, t_i+1, \ldots, t_f$, where $t_i$ and $t_f$ are respectively the initial and final trading (rebalancing) times, being the number of trading steps $N_t = t_f-t_i+1$. We also define $\mu_t$, assets' forecast returns at time $t$, and $\Sigma_t$, the assets' covariance at time $t$. $\mu_t$ is a $N$-length vector while $\Sigma_t$ is and $N \times N$ matrix. For a given trading trajectory (i.e., a given set of vectors $\{\widebar{\w}_{t_i}, \dots, \widebar{\w}_{t_f}\}$), the overall return is given by 
\beq
{\rm Return}  \equiv \sum_{t=t_i}^{t_f} \mu_t^T \widebar{\w}_t, 
\label{retone}
\eeq
and the risk of the trajectory is defined as 
\beq
\label{risky}
{\rm Risk} \equiv \frac{1}{2} \sum_{t=t_i}^{t_f} \widebar{\w}_t^T \Sigma_t \widebar{\w}_t. 
\eeq
Notice that $\widebar{\w}_t^T \Sigma_t \widebar{\w}_t$ is the variance of the portfolio return at time $t$. In practical situations, one typically measures risk at time $t$ in terms of the \emph{volatility}, which is the square-root of this variance. Nevertheless, Eq.(\ref{risky}) is a convenient way to quantify the risk in the dynamic setting since it makes the optimization problem better behaved, being the prefactor $1/2$ a convention.  

The goal of modern portfolio theory is to find the trajectory which maximizes returns for a fixed risk. It is common to request the total investment at any given time is fixed, i.e., 
\beq
 \label{eq:tot_asset_constraint}
 \sum_{n=1}^N \widebar{\w}_{n,t}= K ~~~\forall t,
\eeq
with $K$ the total investment \footnote{An obvious but non-trivial remark: imposing $\sum_{n=1}^N \widebar{\w}_{n,t} \le K$ instead, may actually produce portfolios with higher returns. Notice also that $K$ is the investment, and \emph{not} the value of the investment, e.g. $K=3$ apples, not the value (e.g. in dollars) of those $3$ apples.}. Let us define at this step the normalized weigths
\beq
\w_{n,t} \equiv \frac{\widebar{\w}_{n,t}}{K},  
\eeq
so that their sum at every time $t$ is equal to one. In terms of these normalized weigths, the problem can be solved by finding the trajectory $\{\w_{t_i}, \dots, \w_{t_f}\}$ which minimizes: 
\beq
\label{eq:returns_1}
\mathcal{H} =  \sum_{t=t_i}^{t_f} 
 - \mu_t^T \w_t
 + \frac{\gamma}{2} \w_t^T \Sigma_t \w_t
+  \rho \left(u^T \w_t - 1 \right)^2  . 
\eeq
By analogy to quantum mechanics, we shall refer to the above cost function $\mathcal{H}$ as \emph{Hamiltonian}. $\gamma$ is the \emph{risk aversion}, which tunes the eagerness of the investor to explore risky trajectories, and $\rho$ is a Lagrange multiplier that imposes the constraint in Eq.\,(\ref{eq:tot_asset_constraint}) as a penalty term. We have introduced the $N$-dimensional vector $u$, with $u_n = 1 ~ \forall n$, which makes the constraint in Eq.\,(\ref{eq:tot_asset_constraint}) more compact. The Lagrange multiplier $\rho$ is fine-tuned in order to satisfy Eq.\,(\ref{eq:tot_asset_constraint}).

Note that Eq.\,(\ref{eq:returns_1}) can also be written as 
\beq
\label{eq_sep}
\mathcal{H} = \sum_{t=t_i}^{t_f} h_t, 
\eeq
with 
\beq
h_t \equiv - \mu_t^T \w_t
 + \frac{\gamma}{2} \w_t^T \Sigma_t \w_t
+  \rho \left(u^T \w_t - 1 \right)^2.
\eeq
Thus, the Hamiltonian is \emph{diagonal} in time. This implies that the optimal trading trajectory is simply \emph{the concatenation of the optimal portfolios at each time $t$}. As we will see in the following, if this is the case, then the problem can be solved analytically (in the continuous variable limit). This is not the case when the objective function has terms correlating different times with each other.

Note that the $\w_{n,t}$ are interpreted as a \emph{percentage} of the total investment. 
For instance, having $\w_{n,t} = 0.1$ means that we invest a $10\%$ of our total amount $K$ in asset $n$ at time $t$. Additionally, one may also introduce a cap $K'$ on the maximum amount that can be invested for each asset, i.e., $\w_{n,t} \le K'/K$.

We measure the quality of a portfolio by the so-called \emph{Sharpe ratio}.
\beq
{\rm Sharpe} \equiv \frac{\sum_{t=t_i}^{t_f} \mu_t^T \w_t}{\sqrt{\sum_{t=t_i}^{t_f} \w_t^T \Sigma_t \w_t}},
\eeq
This quantifies the amount of return per unit of risk of trading trajectory. Notice that the numerator is the Return, and not the Profit, which would imply  removing all possible additional costs (such as transaction costs, to be discussed later). Notice also that for e.g. one rebalancing step, the denominator is the normalized volatility, understood as the square root of the variance of the normalized returns. A large Sharpe ratio means a large return for the risk that is assumed, whereas a ratio close to zero means the opposite, and a negative ratio means losses instead of profits.

\subsubsection{Transaction costs}

The problem stated above is simplistic as it does not account for transaction costs. These frequently are comparable to the profit incurred by a given portfolio. The transaction costs are typically a percentage of the transaction (whether buying or selling assets). They can be expressed in terms of unnormalized weigths as:
\beq
\label{trans}
{\rm Cost} \equiv  \sum_{t = t_i}^{t_f} \nu_t |\Delta \widebar{\w}_t| =  \sum_{t = t_i}^{t_f} \sum_{n = 1}^{N}\nu_{nt} |\widebar{\w}_{n,t+1} - \widebar{\w}_{n,t}|,
\eeq
with $\nu_{nt}$ the cost percentage (e.g., $\nu_{nt} = 0.001$ for 10 Basis Points (BPS), meaning a cost of the $0.1\%$ of the total amount of the transaction). The objective function which accounts for these costs in terms of normalized weigths is given by 
\beq
\label{eq:returns}
\mathcal{H} =  \sum_{t=t_i}^{t_f} 
 - \mu_t^T \w_t
 + \frac{\gamma}{2} \w_t^T \Sigma_t \w_t
+ \nu_t |\Delta \w_t | + \rho \left(u^T \w_t - 1 \right)^2.
\eeq
Note that this is not of the form in Eq.\,(\ref{eq_sep}), since $\Delta \w_t$ correlates times $t$ and $t+1$. 

The (percentual) transaction costs are not polynomial in the variables $\w_{n,t}$ because of the absolute value function. This could limit the applicability of some quantum optimization methods. To get around this problem, we can Taylor-expand the absolute value, in a way similar to the expansion in Ref.\cite{crash1} for a step-function. Alternatively, we could introduce ancillary qubits to treat this problem exactly, as explained in Sec.\,\ref{next}.  Here, we choose to approximate Eq.\,(\ref{trans}) by a parabola in the considered range of $\w_{n,t}$:
\beq
\nu_t |\Delta \w_t| \approx \Delta \w_t^T \Lambda_t   \Delta \w_t, ~~ \w_{n,t} \in \left[0, \frac{K'}{K} \right],
\label{trons}
\eeq
with $\Lambda_t$ the best \emph{matrix} of transaction costs at time $t$ that is compatible and realistic with market conditions. This is an excellent approximation to Eq.\,(\ref{trans}) when the $\w_{n,t}$ are discrete variables and $K'/K$ is small. The cost function therefore reduces to:
\beq
\label{eq:returnsfinal}
\mathcal{H} =  \sum_{t=t_i}^{t_f} 
 -\mu_t^T \w_t
 + \frac{\gamma}{2} \w_t^T \Sigma_t \w_t
+ \lambda (\Delta \w_t)^2 + \rho \left(u^T \w_t - 1 \right)^2 , 
\eeq
with $\lambda$ the optimal parabolic coefficient for the transaction costs, as discussed above. Finally, let us remark that in this setting, the percentual profits of the trading trajectory are given by the expression
\beq
{\rm Profit} \equiv \sum_{t=t_i}^{t_f} 
 \left( \mu_t^T \w_t - \lambda (\Delta \w_t)^2 \right), 
\eeq
i.e., the percentual returns minus the percentual costs.

\subsection{Continuous versus discrete formulations}

In this article, we chose to discuss the portfolio optimization problem with discrete variables, which is more relevant to big industry players, as investment funds typically trade in large, discrete amounts. In this section, we will briefly discuss the case where the asset allocations $\w_{n,t}$ can be approximated by continuous variables. This problem is comparatively simpler, as it gives us access to the full toolbox of differential calculus. 

\subsubsection{Continuous asset allocations}

When no transaction costs are present, the problem can actually be solved \emph{exactly}. The minimization of Eq.\,(\ref{eq_sep}) can then be written as
\beq
\frac{\partial  \mathcal{H}}{\partial \w_t^T}= \frac{\partial h_t }{\partial \w_t^T} = 0 ~~~ \forall t 
\eeq
which, using $\mu_t^T \w_t = (\mu_t^T \w_t + \w_t^T \mu_t)/2$ \footnote{A different splitting amounts to a rescaling of the optimal Lagrange multiplier $\rho$, without changing the actual solution.}, amounts to 
\beq
-\frac{1}{2} \mu_t - \rho  u + \left( \frac{\gamma}{2} \Sigma_t + \rho u \cdot u^T \right) \w_t = 0. 
\eeq
Therefore, the optimal solution at time $t$ is given by 
\beq
\label{exactsol}
\w_t = \left(\frac{\gamma}{2} \Sigma_t + \rho u \cdot u^T \right)^{-1} \left( \frac{1}{2} \mu_t + \rho  u \right),  
\eeq
and the optimal dynamic portfolio is just the concatenation of the optimal portfolios at each time $t$. For completeness, one could also get an equation for the multiplier $\rho$:
\beq
\frac{\partial \mathcal{H}}{\partial \rho} = \frac{\partial h_t }{\partial \rho} = 0 ~~~ \forall t, 
\eeq
which in the end simply implies: 
\beq 
\label{cond}
u^T \w_t = 1.
\eeq
This is nothing but the condition in Eq.\,(\ref{eq:tot_asset_constraint}), implying:
\beq
u^T \left(\frac{\gamma}{2} \Sigma_t + \rho u \cdot u^T \right)^{-1} \left( \frac{1}{2} \mu_t + \rho  u \right) = 1, 
\eeq
which is indeed an equation for $\rho$ at each time $t$. In practice, though, it is much easier to simply use Eq.\,(\ref{exactsol}) for a sufficiently large $\rho$, and check a posteriori that Eq.\,(\ref{eq:tot_asset_constraint}) is satisfied up to some degree of accuracy. 

When transaction costs are included, it is no longer possible to solve the problem analytically. In the continuous-time limit, we can recast the problem as a set of \ro{nonlinearly coupled ordinary differential equations}. To do so, we notice that in the discrete formulation the increment in time $\Delta t$ between two consecutive time steps is $\Delta t = 1$. The continuous-time limit can then be taken as $\Delta t \rightarrow 0$, implying  
\beq
\sum_{t=t_i}^{t_f} \Delta t \rightarrow \int_{t_i}^{t_f} dt, ~~~~~~ \frac{\Delta \w_t}{\Delta t} \rightarrow \dot \w, 
\eeq
with $\dot \w$ the \emph{time derivative} of the vector of asset allocations $\w$, which is now a vector field. In this limit, the cost function is given by
\beq
\mathcal{S} = \int_{t_i}^{t_f} L(\w, \dot \w, t)  ~dt.
\eeq
We will refer to $\mathcal{S}$ as the \emph{action} and $L(\w, \dot \w, t)$ as the \emph{Lagrangian}, by analogy to physics. The Lagrangian is obtained by taking the continuous-time limit of the cost function $\mathcal{H}$, Eq.\,\eqref{eq:returnsfinal}. Finding the optimal trading trajectory thus reduces to a conventional functional minimization problem. Our goal is to find the time-path of $\w$ which minimizes the action $\mathcal{S}$:
\beq
\frac{\delta \mathcal{S}}{\delta \w} = 0. 
\eeq
As is well-known, the solution to this equation corresponds to the Euler-Lagrange equations for the Lagrangian, i.e., 
\beq
\frac{\partial L}{\partial \w} - \frac{d}{dt}\frac{\partial L}{\partial \dot \w} = 0. 
\eeq
For a specific problem at hand, it is easy to write the Lagrangian and unfold the above equation, resulting in a set of \ro{nonlinearly coupled ordinary differential equations} for the asset allocations $\w_n(t)$ with initial conditions at $t=t_i$. In this limit, one can thus solve the dynamic portfolio optimization problem using the wide variety of algorithms for systems of \ro{coupled differential equations}. As an example, for Eq.(\ref{eq:returnsfinal}) the Lagrangian is given by 
\beq
L(\w, \dot \w, t) =  -\mu^T \w
 + \frac{\gamma}{2} \w^T \Sigma ~ \w
+ \lambda (\dot \w)^2 + \rho \left(u^T \w - 1 \right)^2, 
\eeq
with $\w, \dot \w, \mu$ and $\Sigma$ being time-dependent. 

\subsubsection{Discrete asset allocations}

In industry, the rebalancing is done at discrete time steps $t$, and it is common for funds to trade assets in large, discrete packages. 

In this setting, the problem is naturally recast as a Quadratic Unconstrained Binary Optimization (QUBO). For this, we choose a binary encoding of each variable $\w_{n,t}$ in terms of $N_q$ bits $x_{n,t,q}$. There are several options for this encoding, as discussed, e.g., in Ref.\cite{Rosenberg2016}. For simplicity in this paper we choose to work with the binary encoding 
\begin{equation}
 \label{eq:encoding}
 \w_{n,t}=\frac{1}{K} \sum_{q=0}^{N_q-1} 2^q x_{n,t,q},
\end{equation}
where $x_{n,t,q} = 0,1$. By construction, we have that the maximum investment per asset is $K' = 2^{N_q}-1$, which is naturally included in the formalism. Investments go also in discrete packages of amount $1$.  Substituting Eq.\,\eqref{eq:encoding} into, e.g., \eqref{eq:returnsfinal} results in a QUBO problem for the $N_{tot} = N \times  N_t  \times N_q$ bit variables, i.e., finding the optimal portfolio weigths at any given time is therefore equivalent to finding the ground state (i.e. the minimum over the variables $\{ x_{n,t,q} \}$) of the classical Hamiltonian 
\begin{equation}
\label{eq:qubo}
 \mathcal{H} = x^T Q x   
\end{equation}
with $x\in \{0, 1\}^{N_{tot}}$ the bit-vector and $Q\in \mathbb{R}^{N_{tot} \times N_{tot}}$ the corresponding QUBO matrix, which can be easily derived from Eqs.(\ref{eq:returnsfinal},\ref{eq:encoding}). 
To solve this problem on a quantum computer, we quantize Eq.~(\ref{eq:qubo}) by promoting the bit variables $\{ x_{n,t,q} \}$ to qubit operators $\{\hat{x}_{n,t,q} \}$ with eigenstates $\ket{0}$ and $\ket{1}$. \ro{Our conclusion at this step is that the portfolio optimization problem, with discrete investments -- as happen in real life --, is a \emph{natural} problem for quantum and quantum-inspired methods. Its formulation is directly a QUBO, by construction, without the need of further mapping the problem to anything else: we can feed this problem to a quantum or quantum-inspired solver \emph{as is}.}

\subsubsection{Discrete problem complexity}

Note that, by applying the transformation $x_i = (1+s_i)/2$, Eq.\,\eqref{eq:qubo} can be mapped to finding the ground state of an Ising spin glass:
\beq
\mathcal{H} = \sum_{i, j=0}^{N_{tot}-1} J_{ij} s_i s_j,
\eeq
where the $s_i = \pm 1$ are spin variables. The couplings $J_{ij}$ can be derived from Eq.\,(\ref{eq:qubo}).

Ising spin-glasses are known to be NP-Hard \ro{in the generic case} \cite{Barahona_1982}, demonstrating \ro{that the portfolio optimization problem can be computationally costly.} This is true even when there are no transaction costs. The optimal trading trajectory is then the concatenation of optimal portfolios at each time step $t$. Finding the optimal trajectory there means solving $N_t$ independent optimization problems for $N \times N_q$ bits each. The instantaneous problem is itself \ro{an Ising spin glass which, in the worst case, may correspond to a hard instance}  (very much unlike in the continuous formulation, which is exactly solvable!).

\section{Methods overview} 
\label{tech}

In this work we use a variety of methods and hardware implementations to solve the dynamic portfolio optimization problem in its discrete formulation. These are the following: 
\begin{enumerate}
\item{Classical: Gekko solver, exhaustive solver.}
\item{Quantum annealing: D-Wave Hybrid.}
\item{Quantum universal: VQE, VQE Constrained.}
\item{Quantum inspired: TN solver.}
\end{enumerate}
The classical methods were implemented as a benchmark of the rest of the algorithms. Gekko is a library which offers tools for non-convex, integer optimization problems \cite{Beal2018}. The exhaustive solver is a brute-force search over valid configurations of the minimum of the cost function. Let us now make a brief overview of the rest of the methods. 

\subsection{D-Wave Hybrid} 
As is well-known, quantum annealing is a type of quantum algorithm based on the ideas of adiabatic quantum computation \cite{aqc}, and is particularly well-suited to solve optimization problems. This process is similar to classical or simulated annealing, where thermal fluctuations allow the system to jump between different local minima in the energy landscape. In quantum annealing, the jumps are mainly driven by quantum tunneling events, which allow for a more efficient exploration of the landscape of local minima, especially when the energy barriers are tall and narrow. 

In this work we used the quantum annealer provided by D-Wave, in particular the so-called D-Wave 2000Q processor, which gives access to 2048 non-coherent qubits coupled through the so-called chimera graph. This architecture allows us to solve problems for up to 65 fully-connected qubits, due to embedding overheads. The D-Wave Hybrid algorithm uses a hybrid classical-quantum strategy to overcomes this limitation, allowing us to deal with much larger problems. In a nutshell, D-Wave Hybrid breaks down problems which are larger than the capability of the quantum processor into parts. These are subsequently recombined to produce the solution.   

\subsection{Variational Quantum Eigensolver}

The Variational Quantum Eigensolver (VQE) \cite{VQE} is a hybrid quantum-classical algorithm for optimization. The idea is to do a variational optimization of a quantum state in order to obtain an approximation to the ground state of a Hamiltonian. This idea is quite generic, but the point of the VQE algorithm is that the ansatz quantum state is, itself, a real quantum state that is implemented on a quantum processor by some quantum circuit. The gates in the circuit depend on parameters, which are the variational parameters of the algorithm. After estimating the energy of the quantum state via sampling, the parameters are then fine-tuned to lower the energy (using, e.g., conjugate gradient). After a number of iterations, the energy converges, producing an approximation to the desired ground state. The performance of VQE depends strongly on several aspects, but most importantly on the choice of variational quantum circuit. In some cases, the search for some complex ground states require of a strongly-entangling quantum circuit, in turn increasing the complexity of the algorithm. However, VQE is still a good option as an optimization tool in current Noisy Intermediate-Scale Quantum (NISQ) processors \cite{NISQ}. 

\begin{figure*}
\centering
\includegraphics[width=\linewidth]{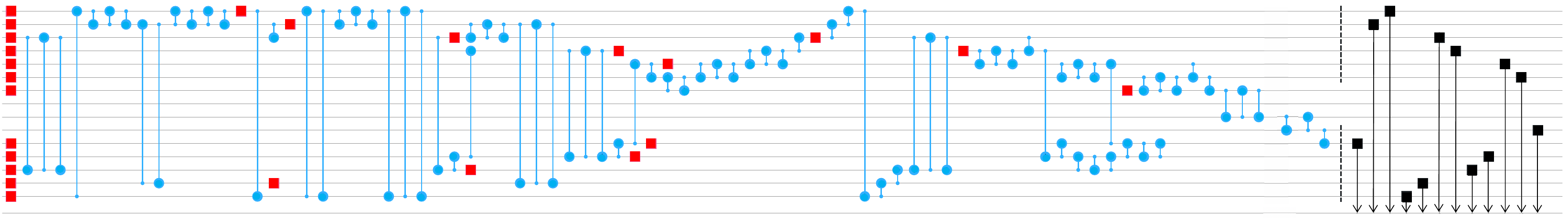}
\caption{(color online) Quantum circuit used for VQE and VQE-Constrained algorithms. The system had 15 qubits on the whole, of which 12 were used in the ansatz. The initial state was $\ket{0}^{\otimes 12}$. In the figure, single-site boxes are one-qubit $y$-rotations, and the two-qubit gates are CNOTs (target qubit being the large circle). In the end, measurements in the computational basis (boxes after the vertical dotted line) are performed on each qubit and stored in a classical register (as shown by the arrows).}
\label{fig:circuit}
\end{figure*}

\ro{For the sake of this paper, we implemented VQE optimization in the quantum processors of IBM-Q, up to 12 qubits. The circuit ansatz for VQE optimization is shown in Fig.\,\ref{fig:circuit}. It is based on a strategy of strongly entangling layers, inspired by the circuit-centric classifier design from Ref.\cite{layers}, with single-qubit rotations around the $y$-axis. In particular, we developed our own VQE algorithm using Xanadu's \emph{PennyLane} library for quantum machine learning \cite{pennylane}, which is well-suited to run on IBM's quantum backend. Our variational quantum circuit consisted of 82 C-NOTs and 24 variational one-qubit rotations, as shown in Fig.\,\ref{fig:circuit}.}  

Moreover, in order to tackle slightly-larger problems than those reachable by plain-vanilla VQE, we implemented a \emph{new and original} approach which we call \emph{VQE-Constrained} for dynamic portfolio optimization. The idea here is quite simple: use VQE to sample several low-energy states at every rebalancing time $t$, and then use a classical approach to find which combination of these provides the highest returns over the whole trading time period. This approach is inspired by low-energy-subspace methods in physics. It follows the intuition that the optimal portfolio can be built in most cases from a combination of near-optimal states. Thus, we identify these states using VQE (this is the computationally expensive part), and then recombine them and estimate their associated profits classically (which is computationally cheap). \ro{Our strategy here was to look for the best 10 solutions at each trading step using VQE. Then, in the post-processing, we tried all possible combinations until finding the one that mininized the cost function.} 

\begin{figure}
\centerline{\includegraphics[width=0.8\columnwidth]{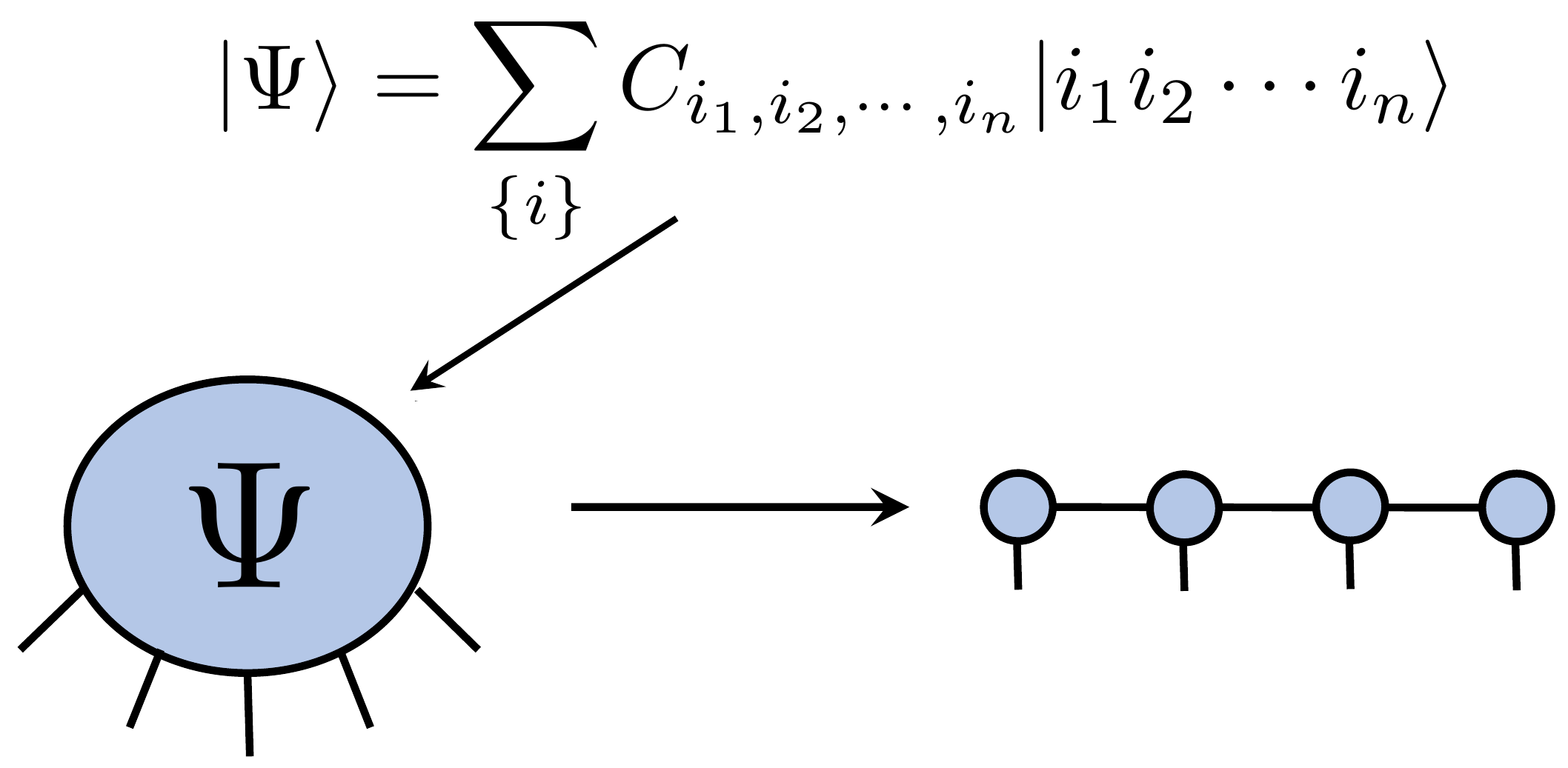}}
\caption{The coefficient of the quantum state of $n$ qubits is a tensor with exponentially many coefficients in the system's size. The inner structure of this tensor is that of a tensor network, which is a network of tensors interconnected by ancillary indices that take into account the structure and amount of entanglement in the quantum state. We represent this here using diagrams, where shapes correspond to tensors, lines to indices, and lines connecting shapes to contracted (summed) common indices. The tensor network on the right hand side is an example of Matrix Product State. \ro{``Open" -- uncontracted -- lines in the tensor network correspond to the degrees of freedom of the original physical qubit (for the case at hand, one line per qubit).}}
\label{fig0}
\end{figure}

\begin{figure*}
\centering
\includegraphics[width=0.8\linewidth]{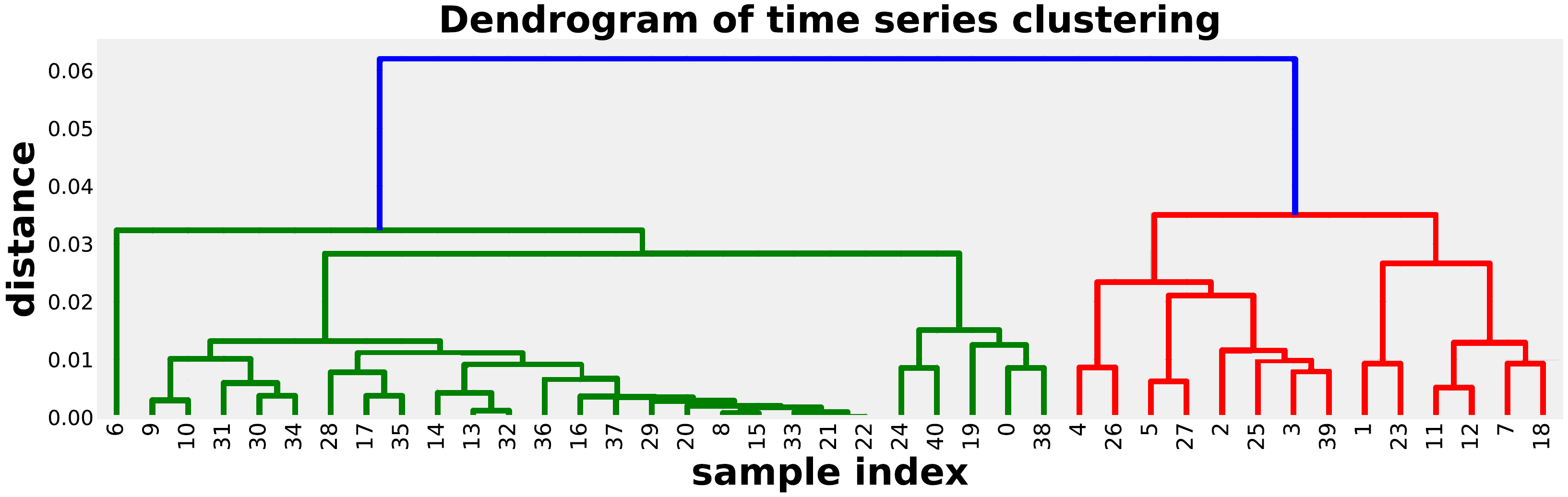}
\caption{Dendogram showing possible data clusters. The Euclidean distance between the different time series is shown on the vertical axis.}
\label{fig:dendogram}
\end{figure*}

\subsection{Tensor Networks}
TNs are representations of complex quantum states based on their local entanglement structure \cite{TN1, TN2}. Take for instance a system of $n$ qubits. Any wave-function of the system can be described inefficiently just by giving its $O(2^n)$ coefficients in the computational basis. As such, these coefficients can be understood as a tensor with $n$ indices, where each index takes two possible values (say, 0 and 1). We could then think of replacing this huge, nasty tensor, by a network of interconnected tensors with less coefficients, see Fig.\,\ref{fig0} for an example. This construction defines a TN, \ro{with each subsystem in the figure corresponding in practice to the Hilbert space of a qubit. Constructed in this way, the TN} depends on $O({\rm poly}(n))$ parameters only, assuming that the rank of the interconnecting indices is upper-bounded by a parameter $D$, which is called  ``bond dimension". Similarly, interconnecting indices in the network are also called ``bond indices", and provide the structure of the many-body entanglement in the quantum state. Any $D > 1$ provides an entangled quantum state. 

As is well-known in physics, TNs are a natural tool to solve optimization problems. People have been using them as an ansatz to approximate low-energy eigenstates of Hamiltonians, and many algorithms have been invented to this aim (see e.g. Ref.\cite{TN2} and references therein). The idea here is that, by mapping optimization problems to Hamiltonian eigenvalue problems, as done in quantum annealing, we can then use the huge machinery of TN techniques and algorithms to solve the optimization problem at hand. 

In our case, we implemented an optimization strategy over the so-called Matrix Product States (MPS) \cite{MPS}. This family of states has been tested already in a variety of algorithms for many physical examples. \ro{For practical reasons, here we use imaginary-time evolution as optimization method}. Moreover, in order to improve the performance, we also tailored \ro{the actual implementation of our optimization} to the specifics of our problem. 

\section{Data preparation}
\label{data}

As a first step, we benchmarked our different algorithms for the optimization problem using random data. Studying real data proved to be more challenging, given the shear size of the dataset and necessity to extract the data trend. Our dataset consists of the daily values of 52 assets over 8 years. Among these assets were government bonds, variable income securities, fixed income securities, etc. 

The bare return for each asset is given by:
\beq
\mu^{bare}_{n,t} \equiv \frac{P_{n,t} - P_{n,t-1}}{P_{n,t-1}}, 
\label{ret}
\eeq
with $P_{n,t}$ the price at time $t$ of asset $n$. Importantly, mathematical expressions often call for the logarithmic returns instead of the bare returns. These are defined as 
\beq
\mu_{n,t} \equiv \log \left( \frac{P_{n,t}}{P_{n,t-1}} \right) = \log \left (1 + \mu^{bare}_{n,t} \right).
\label{logret}
\eeq
There are several reasons why logarithmic returns are preferred over the bare returns \footnote{See, e.g., \href{https://en.wikipedia.org/wiki/Rate_of_return}{https://en.wikipedia.org/wiki/Rate$\_$of$\_$return}.}. In particular they follow a normal distribution, but most importantly for us, \emph{they are time-additive}, and hence justify the sum in Eq.(\ref{retone}). In trading situations, though, the bare returns are usually small ($\mu^{bare}_{n,t} \ll 1$), implying that bare and logarithmic returns are interchangeable at the expense of a very small error, since  
\beq
\mu_{n,t} =  \log \left (1 + \mu^{bare}_{n,t} \right) \approx \mu^{bare}_{n,t} + \Theta\left( (\mu^{bare}_{n,t})^2 \right).
\label{logret}
\eeq

We used a dimensional reduction technique prior to applying our optimization routine. The motivation for this is two-fold.

First, most algorithms cannot tackle the problem in its full complexity. This is mainly true for the VQE algorithm, where dimensional reduction methods are key to solving the problem despite the processor's limited resources. This contrasts with the tensor networks and D-Wave Hybrid algorithms which can actually find close to optimal trajectories even when all of the problem's variables are used.

Second, we found that even the highest performance optimization algorithms tend to get stuck in local minima when the number of variables is truly gigantic. Dimensional reduction methods such as clustering serve the purpose of discarding irrelevant degrees of freedom early. 

We have observed that assets which simultaneously present low variance and low returns tend to not be part of the optimal portfolio. Indeed, investing in these assets promises consistently low returns. By discarding assets which simultaneously have sub-average variance and sub-average returns, our dataset can be reduced from 52 down to 28 relevant assets. 

\begin{figure}
\centering
\includegraphics[width=\linewidth]{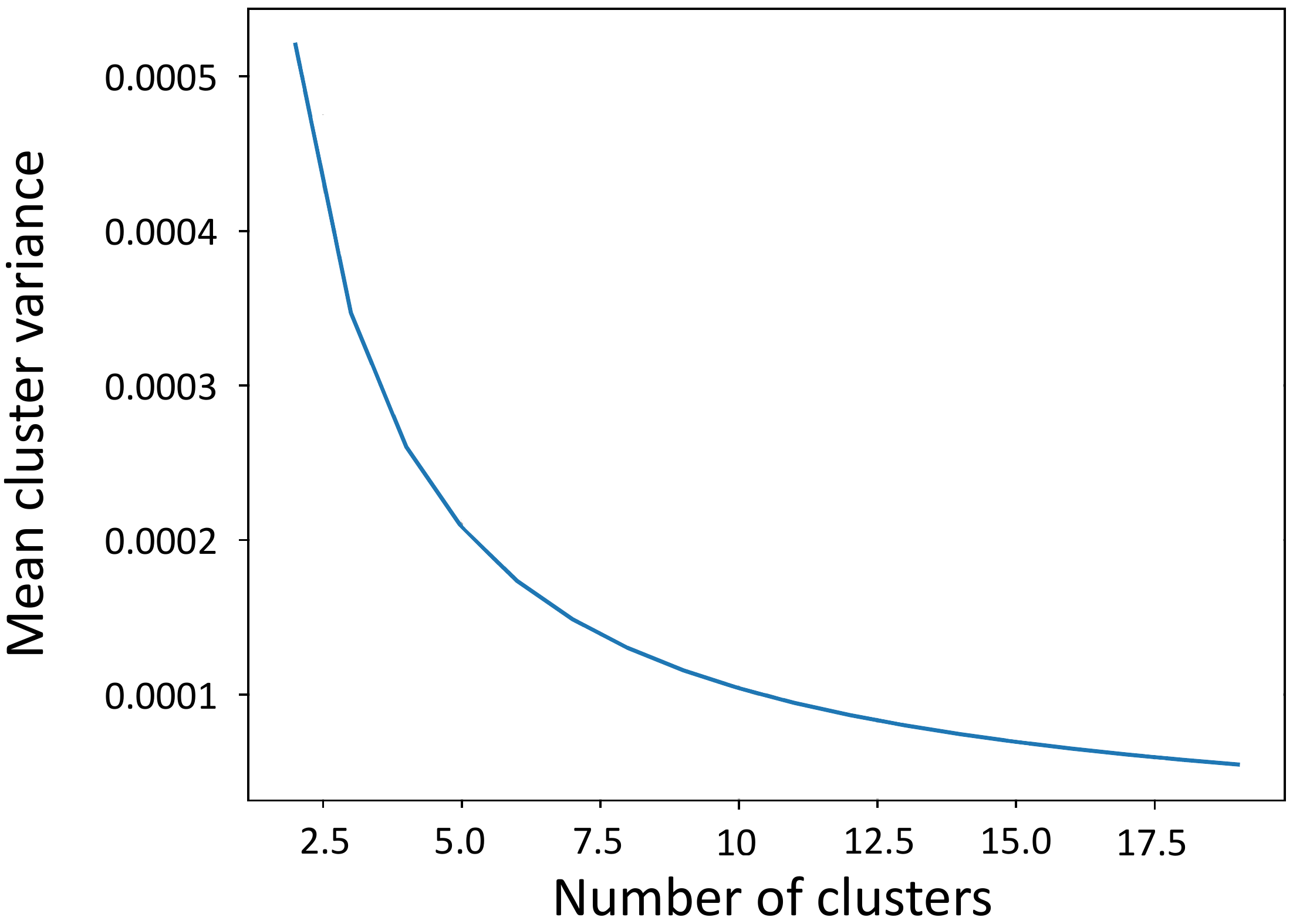}
\caption{Mean variance in each cluster versus the number of clusters.}
\label{fig:elbow}
\end{figure}

\begin{figure}
\centering
\includegraphics[width=\linewidth]{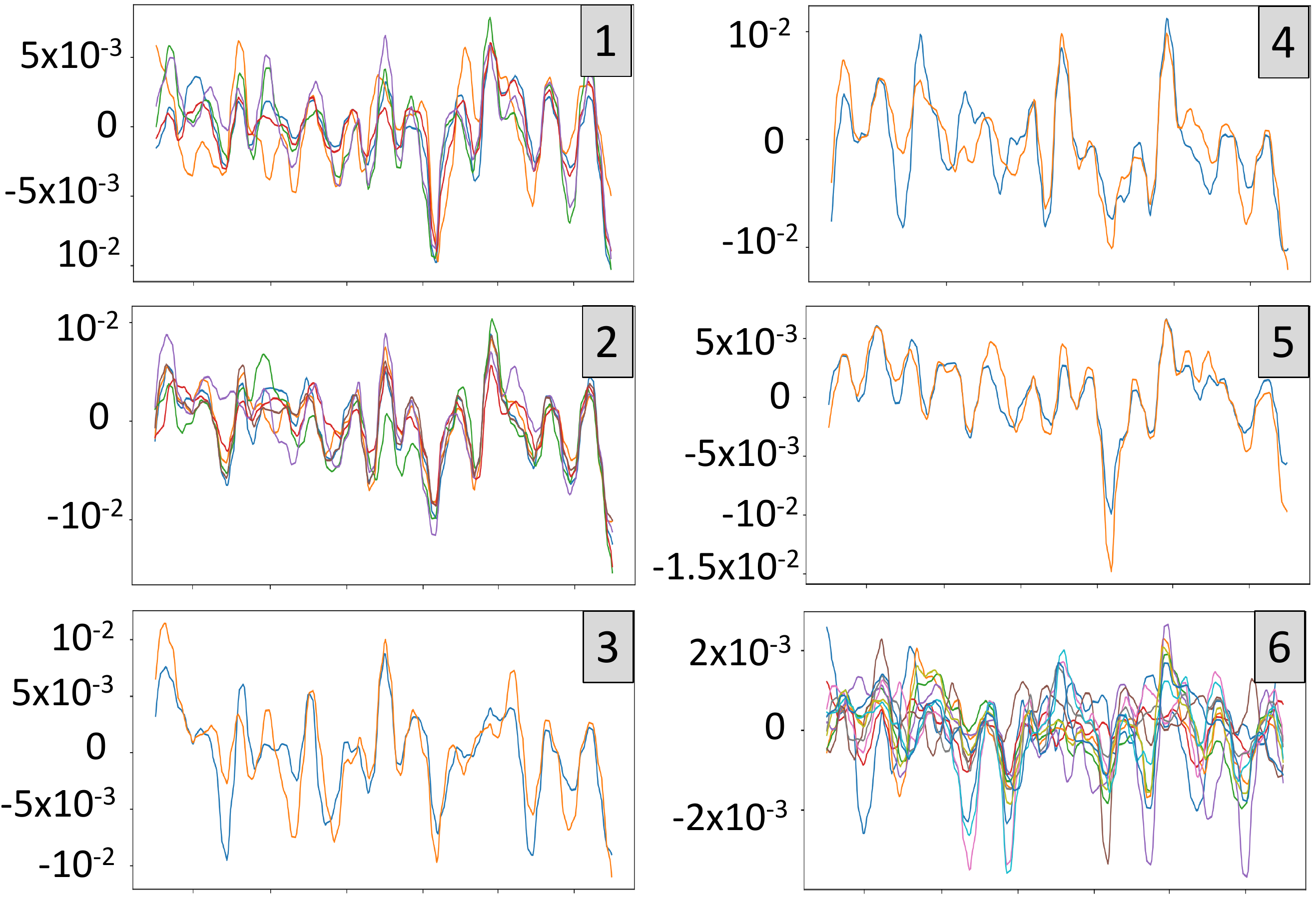}
\caption{Effective assets after clustering, from 52 down to 6. Horizontal axis is the period of time considered for this clustering (two years).}
\label{fig:cluster_trends}
\end{figure}

Prior to clustering, we may also eliminate the noise in the data by applying a Hodrick-Prescott smoothing, which extracts the data trend. We then compute the Euclidean distance in the data trend for each pair of assets. This allows us to identify the degree of correlation between assets, which is represented as a dendogram in the diagram of Fig.\,\ref{fig:dendogram}. To select the optimal number of clusters we wish to consider, we plot the mean cluster variance versus the number of clusters (Fig.\,\ref{fig:elbow}). We can clearly see that, beyond 8 clusters, increasing the number of clusters no longer significantly reduces the mean cluster variance. We therefore take $\approx 6 - 8$ as a reasonable choice of number of clusters for this dataset. The assets' trend is shown in Fig.\,\ref{fig:cluster_trends}, grouped by cluster, for the case of 6 clusters. We observe a good agreement in general for the assets within each cluster. Only one of the clusters shows a relatively high variance between assets' trends. We could always address this issue by considering a larger number of clusters. Alternatively, we could also run several optimization rounds with variable numbers of clusters. By analogy to physics, we actually \emph{renormalized} the assets. Within this picture, the dendrogram in  Fig.\,\ref{fig:dendogram} is nothing but the coarse-graining structure. In practice, after the clustering, the optimization is run over a cost function of \emph{clusterized} assets, i.e., at every time step, the mean investment in each cluster $n$ is the component  $\omega_n$ of the investment vector. In this work, we choose to equally distribute the total investment in a cluster among all the assets within the cluster, i.e., investing $10$ on a cluster with $5$ assets means for us an investment of $10/5$ on each asset belonging to the cluster. Let us stress also that we only applied this clustering for specific methods and hardware, namely, for the ones that could not handle the full problem being analyzed (essentially VQE and VQE Constrained, which run on limited hardware).

We will consider the portfolio rebalancing to happen once monthly, for the sake of concreteness. The returns between transactions are calculated as the sum of returns during the course of that month. Note that we could have calculated them on the basis of a rolling average. This has the advantage of eliminating daily fluctuations, which make it easier to extract trends. On the flip side, the asset returns used in comparison would only approximately reflect the actual asset returns.

The daily covariance is an estimate of the daily fluctuations of a group of assets, encoded in the matrix $\Sigma_t$ at time $t$. We estimate the covariance at time $t$ based on the fluctuations in assets' returns over a window of the prior 20 business days. As a remark, notice also that clustered assets have, by construction, low covariance (since similar assets - i.e. those with large covariance - are the ones being clustered!). In practice, this also means having a smaller risk factor in the final cost function. 

Once we get an optimal portfolio trajectory for clustered effective assets, we \emph{unfold} the investment by assuming an equal investent on each asset forming each specific cluster. From this we obtain the daily return on an equal investment in each asset within each cluster, and in turn an overall trading trajectory in the original variables. 

\section{Results}
\label{results}

Let us now discuss the results obtained with the different techniques \footnote{All portfolio optimizers have been developed by the Multiverse Computing\textsuperscript{\textcopyright} team.} To show the capabilities of each solver, we built datasets of different sizes. We considered datasets of size XS, S, M, L, XL and XXL, each of which pushes a solver to its limit. The S dataset, for instance, is the maximal system size which the exhaustive solver could handle. Details of these datasets are shown in Table \ref{tab:sets}. The risk aversion $\gamma$ represents a compromise between risk and returns. By tuning this parameter, we can find the portfolio which promises the highest returns for any fixed risk. The set of most profitable portfolios for all risks is known as the \emph{efficient frontier}. In all simulations, and for the sake of simplicity, we set the risk aversion parameter $\gamma = 1$.

In Table \ref{tab:sharpe} we show the comparison of Sharpe ratios obtained with the different methods and datasets, in Table \ref{tab:profit} we show the profits (i.e., returns minus transaction costs, percentual), and in Table \ref{tab:time} we show an estimation of the computational running time of our simulations. Entries were left blank when a dataset's size exceed a solver's capabilities.

Note that the solution which minimizes the cost function $\mathcal{H}$ does not always extremize the profit, returns, and Sharpe ratio. For dataset M for instance, the TN simulation produces a trajectory which presents higher profits and Sharpe ratio than other solvers, but is further away from the global minima. Another example is dataset L, for which Gekko produces the largest profits, but the largest Sharpe ratio is obtained with TNs instead (see Tables \ref{tab:sharpe} and \ref{tab:profit}). \ro{In fact, the performance of Gekko is quite remarkable, sometimes even better than quantum and quantum-inspired solutions depending on the metric, but unfortunately the method hits a memory wall around $500$ qubits. In this sense, it would be nice to test also the performance of other quantum-inspired solutions such as Fujitsu's Digital Annealing, Microsoft QIO, and Amazon's Alpha-QUBO, as well as popular methods nowadays such as metaheuristics. Some of these techniques provide also a very good performance that would be worth analyzing. We leave this for future studies.} 


Rosenberg et al. were able so study a system of \emph{24 fully connected qubits at most} (see Ref. \cite{Rosenberg2016} Table IV) using the capabilities of D-Wave's processor from 2016. By way of comparison, with today's D-Wave Hybrid as well as with our TN algorithm we optimized up to 1272 fully-connected qubits, and didn't hit the limit of the two approaches. From our experience, these two algorithms could well handle the full dataset of $28$ relevant assets without any clustering and for reasonable periods of time. The algorithms are under continuous development, but this already shows important progress in the performance and variety of portfolio optimization strategies.

Concerning universal quantum processors, we observed good convergence of VQE for this problem in small datasets. As can be seen from Tables \ref{tab:sharpe}, \ref{tab:profit} and \ref{tab:time}, a naive application of universal quantum processors such as IBM-Q still struggle to tackle optimization problems of commercially relevant dimensions. We were able to obtain good performance using appropriate data preparation. Note, however, that universal quantum processors have seen a phenomenal improvement over the past decade, such that we look forward to seeing progress in this area in the coming years.


We have observed that the D-Wave Hybrid approach works well for this problem, better in fact than a simple D-Wave approach (results not shown). Specifically, D-Wave Hybrid allows us to solve problems up to 1272 fully-connected qubits in 171 seconds, which in our opinion is \emph{really} fast. We conclude that hybrid classical-quantum approaches can be already quite useful for quickly finding good-quality solutions to practical optimization problems. For the sake of comparison, the current version of our TN solver took 116833 seconds in solving the same problem running on a common laptop (a MacBook Pro using Matlab), and a pure quantum-annealing approach in D-Wave 2000Q would only have been able to solve up to 65 fully-connected qubits. We observe also that D-Wave hybrid provides an interesting landscape of potential solutions, see for instance Table \ref{tab:solut}, where we show the energies as well as other relevant quantities computed by the algorithm for an optimization of 8 assets over 53 trading steps. As shown in that table, the different figures of merit are in fact different: minimizing the energy does not necessarily e.g. maximize profits.

\begin{table}[t]
\centering
\begin{tabular}{| c || c | c | c | c | c | c | c | c | c | c |} 
\hline
   ~Param.~ & ~~XS~~   & S   & M   & L   & XL   & XXL   \\ 
   \hline
\hline
   $N$   &   3   & 4 &   4   &   8   & 8 &   8   \\ 
   $N_t$   &   2   & 5 &   7   &   17   & 29 &   53   \\ 
   $N_q$  &   1   & 1 &   1   &   2   & 2 &   3   \\ 
   $N_{tot}$   &   6   & 20 &   28   &   272   & 464 &   1272   \\ 
   $2^{N_{tot}}$    &   $64$   & $O(10^6)$ &   $O(10^8)$   &   $O(10^{81})$   & $O(10^{139})$ &   $O(10^{382})$   \\ 
   $K$   &   2   & 3 &   3   &   5   & 10 &   15   \\ 
   $K'$ &   1   & 1 &   1   &   3   & 3 &  7  \\ 
\hline
\end{tabular}
\caption{Specifics of the different datasets used for benchmarking the different algorithms and hardware platforms. Risk aversion for all datasets is fixed to $\gamma = 1$ and transaction costs to $\lambda = 1$. Time steps $N_t$ are measured in business months (i.e., not considering weekends due to closure of some markets).}
\label{tab:sets}
\end{table}
\begin{table}[t]
\centering
\begin{tabular}{| c || c | c | c | c | c | c |} 
\hline
Method & ~~XS~~ & ~~S~~ & ~~M~~ & ~L~ & ~XL~ & ~XXL~ \\
\hline
\hline
VQE&3.59&-&-&-&-&-\\ 
Exhaustive&6.31&8.90&-&-&-&-\\ 
 VQE  Constrained &6.31&6.04&4.81&-&-&-\\ 
Gekko&5.98&8.90&8.39&15.83&20.76&-\\ 
 D-Wave  Hybrid &5.98&8.90&8.39&7.47&9.70&12.16\\ 
 Tensor  Networks &5.98&8.90&9.54&16.36& 15.77 & 15.83\\ 
\hline
\end{tabular}
\caption{Sharpe ratios computed by the different methods for the different datasets and time periods from Table \ref{tab:sets}}
\label{tab:sharpe}
\end{table}
\begin{table}[t]
\centering
\begin{tabular}{| c || c | c | c | c | c | c |} 
\hline
Method & ~XS~ & ~S~ & ~M~ & ~L~ & ~XL~ & ~XXL~ \\
\hline
\hline
VQE& $2.4\,\%$ &-&-&-&-&-\\ 
Exhaustive& $5.1\,\%$ & $13.9\,\%$ &-&-&-&-\\ 
 VQE  Constrained & $5.1\,\%$ & $9.1\,\%$ & $7.1\,\%$ &-&-&-\\ 
Gekko& $5.8\,\%$ & $13.9\,\%$ & $13.6\,\%$ & $54.1\,\%$ & $71.6\,\%$ &-\\ 
 D-Wave  Hybrid & $5.8\,\%$ & $13.9\,\%$ & $13.6\,\%$ & $18.9\,\%$ & $29.3\,\%$ & $67.6\,\%$ \\
Tensor  Networks & $5.8\,\%$ & $13.9\,\%$ & $15.4\,\%$ & $38.2\,\%$ & $39.6\,\%$ & $39.7\,\%$ \\ 
\hline
\end{tabular}
\caption{Profits (percentual) computed by the different methods for the different datasets and time periods from Table \ref{tab:sets}.}
\label{tab:profit}
\end{table}
\begin{table}[t]
\centering
\begin{tabular}{| c || c | c | c | c | c | c |} 
\hline
Method & ~XS~ & ~S~ & ~M~ & ~L~ & ~XL~ & ~XXL~ \\
\hline
\hline
VQE & 278 & - & - & - & - & - \\
Exhaustive & 0.005 & 34 & - & - & - & - \\
VQE Constrained & 123 & 412 & 490 & - & - & - \\
Gekko&24&27&21&221&261&-\\ 
 D-Wave  Hybrid &8&39&19&52&74&171\\ 
Tensor Networks & 0.838 & 51 & 120 & 26649 & 82698 &116833 \\
\hline
\end{tabular}
\caption{Run-times (in seconds) estimated for the different methods for the different datasets from Table \ref{tab:sets}.}
\label{tab:time}
\end{table}

\begin{table}[t]
\centering
\begin{tabular}{| c || c | c | c | c | c | c |} 
\hline
T & ~E~ & ~C~ & ~R~ & ~TC~ & ~P~ & ~SR~ \\
\hline
\hline
1 & ~-0.673~ & ~0.755~ & ~10.566~ & ~6.514~ & ~4.052~ & ~12.161~ \\
2 & ~-0.661~ & ~0.731~ & ~19.347~ & ~6.026~ & ~4.321~ & ~12.101~ \\
3 & ~-0.656~ & ~0.736~ & ~10.248~ & ~5.662~ & ~4.586~ & ~11.939~ \\
4 & ~-0.621~ &  ~0.741~&  ~9.841~& ~7.366~& ~2.475~& ~11.428~ \\
5 & ~-0.620~ &  ~0.793~&  ~9.736~& ~6.066~& ~3.670~& ~10.931~\\
\hline
\end{tabular}
\caption{Example of landscape of computed solutions with D-Wave Hybrid, for an optimization of 8 assets over 53 trading steps, with maximum holdings 15: Computed Trajectory (T), Energy (E), Covariance (C), Returns (R), Transaction Costs (TC), Profit (P), and Sharpe Ratio (SR). The actual trajectory is not given for space reasons, but is available upon request.}
\label{tab:solut}
\end{table}

We notice that our quantum-inspired TN solver tends to approach the problem's global minimum more reliably than D-Wave Hybrid in some cases. In the case of the M, L and XL datasets, for instance, our TN algorithm returns solutions which have a larger Sharpe ratio and/or larger profits. Furthermore, for the XXL dataset the solution could still be further improved by playing with different hyperparameters and fine-tuning the algorithm further, which is currently work in progress. In any case, it is  interesting to notice that whereas D-Wave Hybrid provides the largest profits for the XXL dataset, the largest Sharpe ratio is however given by the TN solver. This deserves a deeper analysis, which we leave for future works.

We anticipate that our TN results can still be largely improved, both in terms of accuracy and performance. Based on our own estimations, we can reduce \ro{at least $\approx 100\times$} the run-time of our TN algorithm (see Sec.\,\ref{next}), as, in its present version, the TN algorithm is not memory intensive. \ro{In fact, acceleration of the TN code is possible at three levels: (i) software, (ii) hardware, and (iii) the algorithm itself. All in all, our} work proves that TN solvers \ro{are} a \emph{cheap and practical} way to tackle large combinatorial optimization problems, compared to other quantum and quantum-inspired approaches. 

\section{Next steps}
\label{next}

Let us now discuss what would be the next steps to make our algorithms better tailored to real-life situations, as well as to improve their efficiency. 

\subsection{More constraints} 

As we said in the introduction, the portfolio optimization problem is central to quantitative finance. We have discussed its dynamic version, where we search for an optimal trading trajectory given the price history of several assets. There exist many commercially relevant variations of this problem. In this section, we will discuss three: the market impact, exact transaction costs, and the $10-5-40$ rule. 

\subsubsection{Market impact}

Intuitively, when a large order is passed, it can affect the market. For instance, a large ``buy" order may increase the price of an asset because it signals high demand, whereas a large ``sell" order may instead decrease it. This, of course, would alter the optimal trading trajectory. As discussed in Ref.\cite{Rosenberg2016} this can be implemented as
\beq
{\rm Market} = \sum_{t = t_i}^{t_f} \Delta \widebar{\w}_t^T \Lambda_t' \widebar{\w}_t,
\label{market}
\eeq
with $\Lambda_t$ a diagonal matrix of market impact coefficients.

Eq.\,(\ref{market}) can be interpreted as follows: the impact of a trade in asset $n$ at time $t$ on the value of our portfolio is proportional to $\Delta \widebar{\w}_{n,t}$, the amount of shares of asset $n$ bought or sold, and to  $\widebar{\w}_{n,t}$, the amount of shares of asset $n$ held at time $t$.

Note that if we are buying asset $n$, $\Delta \widebar{\w}_{n,t} > 0$, we expect the asset's value to increase. Our portfolio's returns should then grow as $\widebar{\w}_{n,t}$. The matrix $\Lambda_t'$ is therefore positive definite. Similarly, when selling, $\Delta \widebar{\w}_{n,t} < 0$, and our portfolio's returns drop with $\widebar{\w}_{n,t}$.

The resulting classical Hamiltonian including market impact is therefore 
\beqa
\label{eq:returnsmarket}
\mathcal{H} =  \sum_{t=t_i}^{t_f} 
 &-&\mu_t^T \w_t
 + \frac{\gamma}{2} \w_t^T \Sigma_t \w_t
+ \lambda (\Delta \w_t)^2 \nonumber \\ 
&-& \Delta \w_t^T \Lambda_t' \w_t + \rho \left(u^T \w_t - 1 \right)^2  ,  
\eeqa
again in terms of normalized weigths. Note that this Hamiltonian is in the form of a QUBO.

As can be inferred from Eq.\,(\ref{market}), this term is non-negligible only when (i) we have a large amount of the asset being traded in our portfolio, and (ii) the traded amount is large.

\subsubsection{Exact linear transaction costs}
As discussed in Sec.\,(\ref{problem}), we would expect the transaction costs to be linear in the absolute value of the transaction magnitude. We approximated these by a parabola, as polynomial expressions find a more natural formualtion as a QUBO. That option is cheap, fast, and works quite well for the accuracies that are normally considered. It is however possible to \emph{exactly} implement the linear transaction costs by introducing $N \times N_t$ extra ancillary qubits $y_{n,t}$. This can be done by noticing that the sign of the difference $(\w_{n,t+1} - \w_{n,t})$ in Eq.\,(\ref{trons}) can be controlled by $y_{n,t}$ as follows: 
\beq
\nu_{nt} |\w_{n,t+1} - \w_{n,t}| \rightarrow  \nu_{nt} (\w_{n,t+1} - \w_{n,t})(1-2y_{n,t}), 
\eeq
with $y_{n,t} = 0$ if $(\w_{n,t+1} - \w_{n,t}) > 0$ and $y_{n,t} = 1$ otherwise. This condition can be further imposed by including the penalty term 
\beq
\rho' \sum_{t=t_i}^{t_f} \sum_{n=1}^N(\w_{n,t+1} - \w_{n,t}) y_{n,t}, 
\eeq
with $\rho'$ a Lagrange multiplier. So, in this way, one could introduce the linear transaction costs in the QUBO formulation (notice that the above equations are quadratic), but at the expense of introducing more qubits, thus increasing the complexity of the optimization problem. 

\subsubsection{$10-5-40$ rule}

It is common for investors to be bound by a constraint known as the $10-5-40$ rule: no investment in an asset can represent more than 10\% of the total portfolio's value; investments larger than $5\%$ of the portfolio's value cannot total $40\%$ of the portfolio's value. In the following we will demonstrate a way to implement this constraint as a QUBO, but at the price of increasing the complexity of the optimization problem.   

If we choose to impose the $10-5-40$ rule, we could proceed as follows. First, choose $K' = 0.1 \times K$ so that variables $\w_{t,n} \in [0,K'/K]$ directly satisfy the $10\%$ constraint, which can be achieved by choosing an appropriate $N_q$. Notice that, in our previous simulations, the ratios $K'/K$ can be extracted from Table \ref{tab:sets}, but we could easily impose $K'/K = 0.1$. Next, we impose that the sum of those weights larger than 5$\%$ does not amount to more than 40$\%$ of the total investment. We can do this by adding a penalty term which uses \emph{Slack variables} $\alpha_t \in {\mathbb R}$. The idea is that, by minimizing also over the Slack variables, one is able to convert an inequality constraint into an equality constraint.  Additionally, we include ancillary qubits $y_{n,t}$ such that $y_{n,t} = 1$ if $\w_{n,t} > 0.05$ and $y_{n,t} = 0$ otherwise. The penalty term for the $40\%$ constraint is given by 
\begin{equation}
 \label{eq:51040_2}
{\rm UpToForty} = \rho' \sum_{t=t_i}^{t_f} \left(\sum_{n=1}^N y_{n,t} \w_{n,t} - 0.4 + \alpha_{t}^2 \right)^2, 
\end{equation}
with $\rho'$ a Lagrange multiplier. Finally, we need to ensure that $y_{n,t} = 1$ if and only if $\w_{n,t} > 0.05$, which can be accounted for by the penalty term 
\beqa
 \label{eq:51040_3}
{\rm WhoIsFive} = &\rho''& \sum_{t=t_i}^{t_f}  \sum_{n=1}^N  y_{n,t}(0.05 - \w_{n,t}  +\mu_{n,t}^2)^2  \nonumber \\ 
&+& (1-y_{n,t})(\w_{n,t} - 0.05 + \nu_{n,t}^2)^2 , 
\eeqa
which again uses a Lagrange multiplier $\rho''$ and Slack variables $\mu_{n,t}, \nu_{n,t} \in {\mathbb R}$.

If we include the $10-5-40$ rule, the optimization problem becomes much more complex because of two reasons: first, Eqs. \eqref{eq:51040_2} and \eqref{eq:51040_3} make this a higher-order optimization problem, since Eq \eqref{eq:51040_2} is quartic and Eq.\,\eqref{eq:51040_3} is cubic in the bit variables. Therefore, the problem is now a HUBO (``H" for ``higher-order") instead of a QUBO. HUBOs are generally more complex than QUBOs, and there exist few computational architectures adapted for solving them (though they do exist). Second, the $10-5-40$ rule is a set of inequality constraints, which involve the inclusion of new hyperparameter Slack variables $\alpha_{t}, \mu_{n,t}$ and $\nu_{n,t}$ which also need to be optimized, thus increasing the complexity of the problem. 

\subsection{Improved hardware and codes}

We believe that it should be possible to handle bigger problems in a number of ways. Concerning quantum processors, D-Wave's processor``Advantage" with \emph{Pegasus} topology is able to cope with a larger number of variables (5000+ qubits, 35000+ couplers, 15 couplers per qubits, 23x more optimal solutions than 2000Q for satisfiability problems and 20x faster), see for instance the numerical benchmarks in D-Wave's documentation \cite{DWavePegasus}. Also, hybrid solutions such as D-Wave Hybrid benefit a lot from such developments. Concerning IBM-Q, we have found that a naive application of VQE was very rather limited for this problem, but obtained promising results by constraining the solution space using VQE. We think that it is worth exploring more hybrid solutions of this type. As a matter of fact, this is also the natural state of affairs: in the NISQ era, hybrid quantum-classical solutions (both in hardware and software) are offering the best operational performances, which is also somehow expected. 

Our TNs code is able to handle $O(10^3)$ qubit variables using a MacBook Pro without problems, even though the performance can be improved in a number of ways. We believe that a highly optimized code in C++ fully parallelized on a HPC cluster should be able to handle \emph{really} large problems very efficiently. GPU computing, \ro{FPGAs}, and even Optical Processing Units (OPUs), should also provide improvements at the hardware level. Moreover, at the algorithmic level, it should be possible to try different TN strategies that should also improve the performance. The combination of all these easily mean \ro{more than $100\times$} acceleration of the calculations. Since memory  \ro{is not} a constraint for our algorithm, this would also mean the ability to simulate much bigger systems than the ones considered in this paper. And of course, this option would still be cheaper than other alternatives. 

\vspace{20pt}

\section{Conclusion and outlook}
\label{conclusion}

In this paper we have solved the problem of dynamic portfolio optimization using a number of quantum and quantum-inspired algorithms on different hardware platforms. We ran our solver on the daily values of 52 assets over 8 years. We implemented classical solvers (Gekko, exhaustive), D-Wave Hybrid quantum annealing, two different approaches based on VQE on IBM-Q (one of them brand-new, which we dubbed ``VQE-Constrained"), and for the first time in this context also a quantum-inspired optimizer based on Tensor Networks. We also implemented a preprocessing based on clustering of assets. From our comparison, we conclude that D-Wave Hybrid and Tensor Networks are able to handle the largest possible systems in the present implementation, i.e., those for the XXL dataset. More specifically, with these two methods we managed to solve optimization problems up to 1272 fully-connected qubits, for demonstrative purposes and without hitting the limiting computational capabilities (we could in fact have targeted larger systems). For comparison, previous quantum dynamic portfolio optimizations involved up to 72 qubits (see Table VI in Ref.\cite{Rosenberg2016}). We observed also that D-Wave Hybrid is remarkably fast, whereas Tensor Networks sometimes provide better portfolios at the expense of a longer calculation time. To the best of our knowledge, our work is the first application of VQE and TNs to solve a dynamic portfolio optimization problem with real data. Moreover, our VQE-Constrained approach is also unique, as far as we know. Our preprocessing of real data in order to fit the current capabilities of quantum processors is also novel in this context.  

From our results we also conclude that there seems to be no clear answer as to which is the ``best" algorithm and hardware platform to solve the dynamic portfolio optimization problem for large systems. This is because there are several figures of merit at play: profits, Sharpe ratio, time cost, and also money cost. The performance of the algorithms is  different depending on the figure of merit, leading us to conclude that, in practice, \emph{the more options we have, the better.}

We have good reasons to think that the results presented in this paper are very promising. They show how real data can be handled by upcoming quantum computers, and also show the potential of quantum-inspired methods such as Tensor Networks. We also realize the importance of hybrid approaches, combining quantum and classical processing. This has been the key to improved results, e.g., in VQE-Constrained and D-Wave Hybrid. In fact, a hybrid approach combining quantum processing and Tensor Networks should be quite successful for many problems. This is a topic that we are currently investigating in the broad sense. We also believe that all these developments, involving quantum and quantum-inspired techniques, will change the way quantitative finance is done, and for good. 

\bigskip
{\bf Acknowledgments.-} We thank Hossein Sadeghi for insight about D-Wave, Mhedi Bozzo-Rey for insight about IBM-Q, as well as Christophe Jurczak, Pedro Luis Uriarte, Pedro Mu{\~n}oz-Baroja, Creative Destruction Lab, BIC-Gipuzkoa, DIPC, Ikerbasque, and Basque Government for constant support. 

\bigskip
{\bf Competing Interests.-} The authors declare that there are no competing interests.

\bigskip
{\bf Data Availability.-} The datasets generated during and/or analysed during the current study are available from the corresponding author on a reasonable request. 

\bigskip
{\bf Code Availability.-} The code used during the current study is available from the corresponding author on a reasonable request. 
 
\bigskip
{\bf Author Contribution.-} All authors contributed to the generation of ideas, algorithm design, algorithm implementation, analysis of results, and writing of the paper.  

\bibliography{bibliography}{}

\end{document}